\documentclass[a4paper,fleqn,usenatbib]{mnras}

\usepackage{newtxtext,newtxmath}
\usepackage{xcolor}

\usepackage[T1]{fontenc}
\usepackage{ae,aecompl}

\usepackage{graphicx}	
\usepackage{amsmath}	
\usepackage{amssymb}	
\usepackage{multirow}
\usepackage{xcolor}
\usepackage{subfigure}

\newcommand{\PGC}{2WHSP~J073326.7+515354}	

\title[2WHSP~J073326.7+515354]{Optical spectral characterization of the the TeV extreme blazar 2WHSP~J073326.7+515354}

\author[Becerra Gonz\'alez]{
J. Becerra Gonz\'alez$^{1,2}$\thanks{E-mail: jbecerra@iac.es},  J. A. Acosta--Pulido$^{2,1}$\thanks{E-mail: jose.acosta@iac.es} and R. Clavero$^{2,1}$
\\
$^{1}$Universidad de La Laguna (ULL), Departamento de Astrof\'isica, E-38206 La Laguna, Tenerife, Spain\\
$^{2}$Instituto de Astrof\'isica de Canarias (IAC), E-38200 La Laguna, Tenerife, Spain
}

\date{Accepted XXX. Received YYY; in original form ZZZ}

\pubyear{2020}
 
\begin{document}
\label{firstpage}
\pagerange{\pageref{firstpage}--\pageref{lastpage}}
\maketitle

\begin{abstract}
The emission from the relativistic jets in blazars usually outshines their host galaxies, challenging the determination of their distances and the characterization of the stellar population. The situation becomes more favorable in the case of the extreme blazars (EHBLs), for which the bulk of the emission of the relativistic jets is emitted at higher energies, unveiling the optical emission from the host galaxy. The distance determination is fundamental for the study of the intrinsic characteristics of the blazars, especially to estimate the intrinsic gamma-ray spectra distorted due to the interaction with the Extragalactic Background Light. In this work we report on the properties of 2WHSP~J073326.7+515354 host galaxy in the optical band, which is one of the few EHBLs detected at TeV energies. We present the first measurement of the distance of the source, $\mathrm{z}=0.06504\pm0.00002$ (velocity dispersion $\sigma=237 \pm 9\,\mathrm{km s^{-1}}$). We also perform a detailed study of the stellar population of its host galaxy. We find that the mass-weighted mean stellar age is $11.72\pm0.06\,\mathrm{Gyr}$ and the mean metallicity $[M/H]=0.159 \pm 0.016$. In addition, a morphological study of the host galaxy is also carried out. The surface brightness distribution is modelled by a composition of a dominant classical bulge ($R_e=3.77\pm1\arcsec $ or equivalently 4.74~kpc) plus an unresolved source which corresponds to the active nucleus. The black hole mass is estimated using both the mass relation with the velocity dispersion and the absolute magnitude from the bulge yielding comparable results: $(4.8\pm0.9)\times10^8\,M_{\odot}$ and $(3.7\pm1.0)\times10^8\,M_{\odot}$, respectively.

\end{abstract}

\begin{keywords}
BL Lacertae objects: general -- BL Lacertae objects: individual (2WHSP~J073326.7+515354) -- galaxies: active -- galaxies: nuclei
\end{keywords}



\section{Introduction}\label{intro}
Radio loud Active Galactic Nuclei (AGN) whose relativistic jets point close to the direction to Earth are generally called blazars. The emission from this type of sources is strongly dominated by their jets because of the beaming caused by geometrical effects. Therefore, the broadband emission from blazars is dominated by the continuum emission of the jet from radio to gamma rays, typically displaying two bumps on their spectral energy distribution (SED). Two classification methods are usually applied to blazars. On one hand, based on their optical spectra, blazars are classified as Flat Spectrum Radio Quasars (FSRQs) if their optical spectra show broad and intense emission lines (equivalent width ($W_{\lambda}$) $>5$\,\AA) and BL Lac objects (BL Lacs) in case their optical spectra is dominated by the continuum emission from the jet, showing faint spectral lines if any. On the other hand, blazars can be classified with respect to the position of the peak from the first bump (synchrotron peak) on their SED. In case of BL Lacs, they are classified as low (LBL, with $\nu_{\text{peak}}^{\text{sync}}<10^{14}$\,Hz), intermediate (IBL, with $10^{14}\leq\nu_{\text{peak}}^{\text{sync}}<10^{15}$\,Hz), high (HBL, $10^{15}\leq\nu_{\text{peak}}^{\text{sync}}\leq10^{17}$ Hz) and extreme (EHBL, $\nu_{\text{peak}}^{\text{sync}}>10^{17}$\,Hz) peaked BL Lacs. \\

2WHSP~J073326.7+515354 (a.k.a. PGC~2402248) is included in the 2WHSP catalog \citep{2WHSP_cat} as an EHBL with \mbox{$\nu_{\text{peak}}^{\text{sync}}=10^{17.9}$~Hz}. Recently, \cite{pgc_magic} reported compatible results,  \mbox{$\nu_{\text{peak}}^{\text{sync}}=10^{17.8 \pm 0.3}$~Hz}. The source is a unknown redshift gamma-ray emitter, detected in the high-energy band (HE, E$>$100 MeV) by {\it Fermi}-LAT \citep[also known as 4FGL~J0733.4+5152,][]{4fgl}. Recently, it was also detected in the very high energy (VHE, E$>$100 GeV) gamma-ray band with the MAGIC telescopes \citep{pgc_magic}, becoming one of the few detected EHBLs at TeV energies. Therefore, 2WHSP~J073326.7+515354 is a source of particular interest to populate the EHBL parameter space at gamma rays, which detectability is very challenging with the current instrumentation. \\

Gamma rays are absorbed due to the interaction with the Extragalactic Background Light (EBL) via pair production \cite[see e.g.][and references therein]{ebl_magic}. Hence, in order to study the intrinsic gamma-ray emission from blazars, the distance is a key parameter to infer the absorption and distortion that the gamma-ray spectra suffer due to the gamma-ray absorption on its way to Earth. Generally, measuring the redshfit from BL Lacs is challenging due to the fact that their optical spectra are typically dominated by continuum emission from their jets, outshining the weaker emission from their host galaxies. This issue is not only affecting the redshift measurements but also the characterization of their host galaxies. Actually, the knowledge of the host galaxies for blazars is very limited and mostly related to low luminosity sources \citep[see e.g.][]{2004A&A...413...91S,2004A&A...413...97G}. However, in case of EHBLs, the SED bumps are located at high energies, allowing us to clearly identify the emission from their host galaxy with a reduced contamination from the jet emission in comparison with other types of blazars. This is the case of 2WHSP~J073326.7+515354, which optical spectrum is dominated by its host galaxy emission. The high quality of the observations obtained with the 10.4\,m Gran Telescopio de Canarias (GTC) allows us to determine the redshift precisely as well as to carry out a detailed analysis of the host galaxy. \\

In Section~\ref{sec:observations} the optical observations and data analysis procedures are presented. The study on the stellar population of the host galaxy is discussed in Section~\ref{sec:StellPop}, and a morphological study of the host galaxy can be found in Section~\ref{sec:morphology}. The results and conclusions of this work are summarized in Section~\ref{sec:conclusions}.

\section{Optical spectroscopy. Observations and data reduction}
\label{sec:observations}

Optical observations were carried out at two epochs from the Roque de los Muchachos observatory located in the canary island of La Palma. A first optical spectrum of the target was taken using the 2.5\,m Isaac Newton Telescope (INT) on 2017 September 21. It allowed us to identify three possible absorption features, estimating a preliminary redshift. After the detection of VHE gamma rays associated with this target with the MAGIC telescopes \citep{pgc_magic}, a second epoch was observed on 2018 May 2 using the GTC to improve the S/N ratio. This second observation is crucial for the detailed study of its stellar population as presented in the following sections.\\

\begin{figure*}
\centering
\includegraphics[width=1.\textwidth]{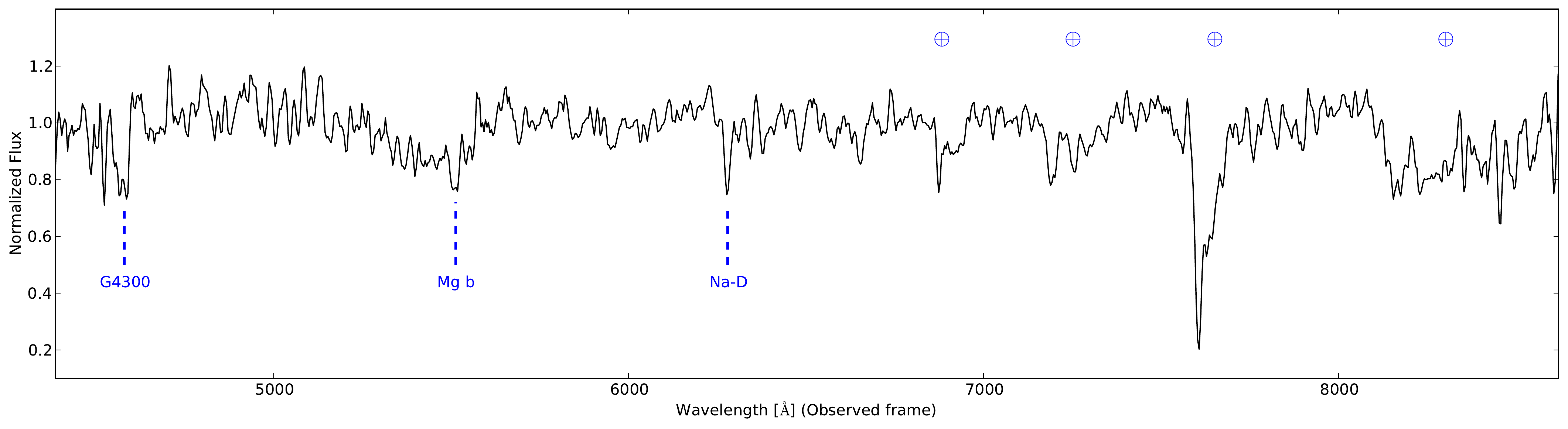}
\caption{Normalized optical spectrum of 2WHSP~J073326.7+515354 as observed with the INT telescope on 2017 September 21. Besides the telluric emission lines, three absorption features are identified in the spectrum. A smoothing of 3 pixels is used for display.}
\label{fig:spectrum_int}
\end{figure*}

The observations with INT were carried out using the long-slit IDS spectrograph, the camera RED+2, the grating R300V and 1\,arcsec slit width, yielding a resolution of 1067 at 5000\,\AA. The slit was oriented along the parallactic angle. The target was observed at an airmass of 1.3, taking 6 exposures of 100\,s each (a total of 600\,s). The data were reduced\footnote{IRAF standard packages were used for the different reduction tasks through this work.} following the standard procedures for bias subtraction and flat-field correction. The spectrum was calibrated in wavelength using a CuAr+CuNe calibration arc using the same instrumental setup as for the observation of the target. The normalized spectrum is shown in Fig.~\ref{fig:spectrum_int}, were three absorption features can be identified: G4300 at 4581\,\AA, Mg b at 5512\,\AA\ and Na-D at 6278\,\AA. The identification of these absorption features are consistent with a redshift of z$=0.0649\pm0.0004$. The error has been estimated by performing Montecarlo simulations of the observed spectrum with the uncertainty estimated from the RMS of the continuum. The absorption features of the simulated spectra are then fitted with a Gaussian function and the error is obtained from the resulting distribution. For this calculation, the Na-D line is used since it is best defined line identified in the spectrum.

The second optical observation of the target was obtained using the 10.4\,m GTC telescope. The observations were carried thanks to a DDT proposal (GTC2018-099). The instrumental setup used OSIRIS in spectroscopic long-slit mode, with the grism R1000B and slit width of 1 arcsec, which yields a spectral resolution of 625 at 5000\,\AA. Three exposures of 300\,s each were taken on the target, resulting on a total of 15\,min observation time. The observations were performed in parallactic angle at an airmass of 1.3. In addition to the target, the calibration star Ross640 was also observed using the same instrumental setup as well as observation conditions including the same airmass. For the calibration star, three observations of 15\,s each were taken. Standard calibration images for bias, flat field and calibration lamp were also taken during the same night.

The GTC data were reduced following the standard procedures for bias subtraction and flat-field correction. Special care was taken to subtract the sky lines averaging the sky spectrum as observed adjacent to the target. The spectrum was extracted using an aperture of 4 arcsec around the emission peak. The spectrum was flux calibrated using an spectrophotometric standard taken after our target using the same configuration. Afterwards, the target spectrum was corrected for telluric absorption using the tool {\it telfit} \citep{Gullikson14}. Finally, the spectrum is corrected from interstellar reddening by a value $A_V=0.165$, taken from NED database which uses \citet{Schlafly11}. The extinction curve provided by \cite{Fitzpatrick07} was used in this work.  

\section{Stellar Population}
\label{sec:StellPop}

In order to characterize the stellar population emission dominating the observed spectrum, we used 
the penalized pixel fitting technique (pPXF) \citep{Cappellari04, Cappellari17}. We adopted the stellar population synthesis models MILES\footnote{Medium--resolution Isaac Newton Telescope Library of Empirical Spectra} \citep{Vazdekis10} as template spectra. These models have a spectral resolution of $\mathrm{FWHM}= 2.5$\,\AA\ ($\sigma_{instr} \sim 64 \mathrm{km}  \mathrm{s}^{-1}$), as estimated in \citet{Falcon11}, and cover the range 3525--7500\,\AA\ \citep{SanchezBlazquez06}. 
We selected the set of Single Stellar Population (SSP) models which uses the Kroupa Universal function \citep{Kroupa01} as initial mass function (IMF),  and the set of isochrones from the Bag of Stellar Tracks and Isochrones models \citep[BaSTI,][]{Pietrinferni13}. The grid of models covers a range of stellar ages from 0.03 to 14\,Gyr and metallicity from [M/H]~-2.27 to 0.4.  

\begin{figure*}
\centering
\includegraphics[width=0.98\textwidth]{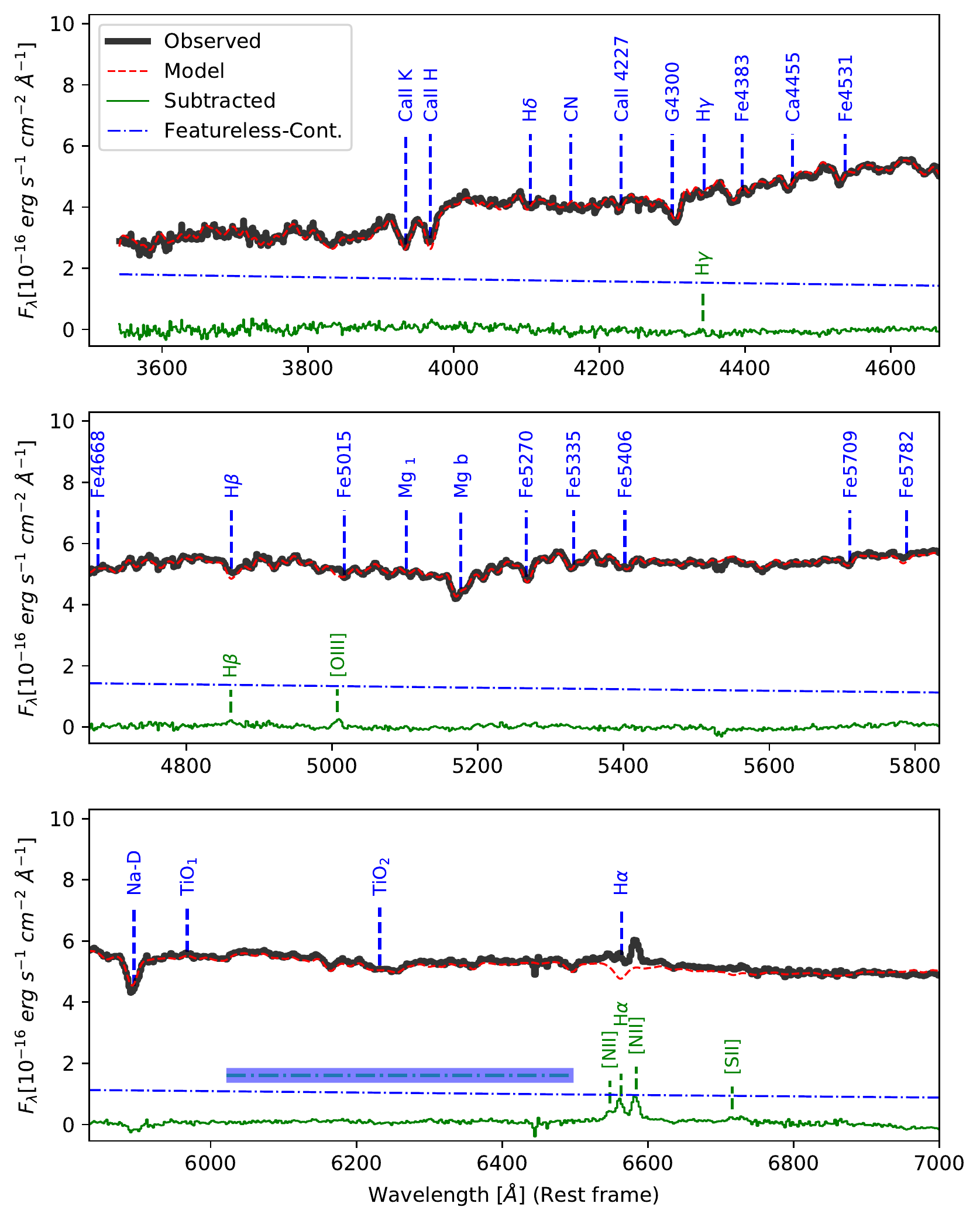}
\caption{Optical spectra of \PGC\ (black line) and the best fit stellar population model (red line). Note the good matching of the stellar population synthesis model to the observed spectrum. The subtraction of the model to the observed spectra is shown as residual (green line). The most conspicuous features are indicated in the figure. The horizontal bar in the bottom panel represent the flux level expected for the unresolved nucleus as derived from the surface brightness model.}
\label{fig:stellarpop}
\end{figure*}

\begin{figure}
    \centering
    \includegraphics[width=0.48\textwidth]{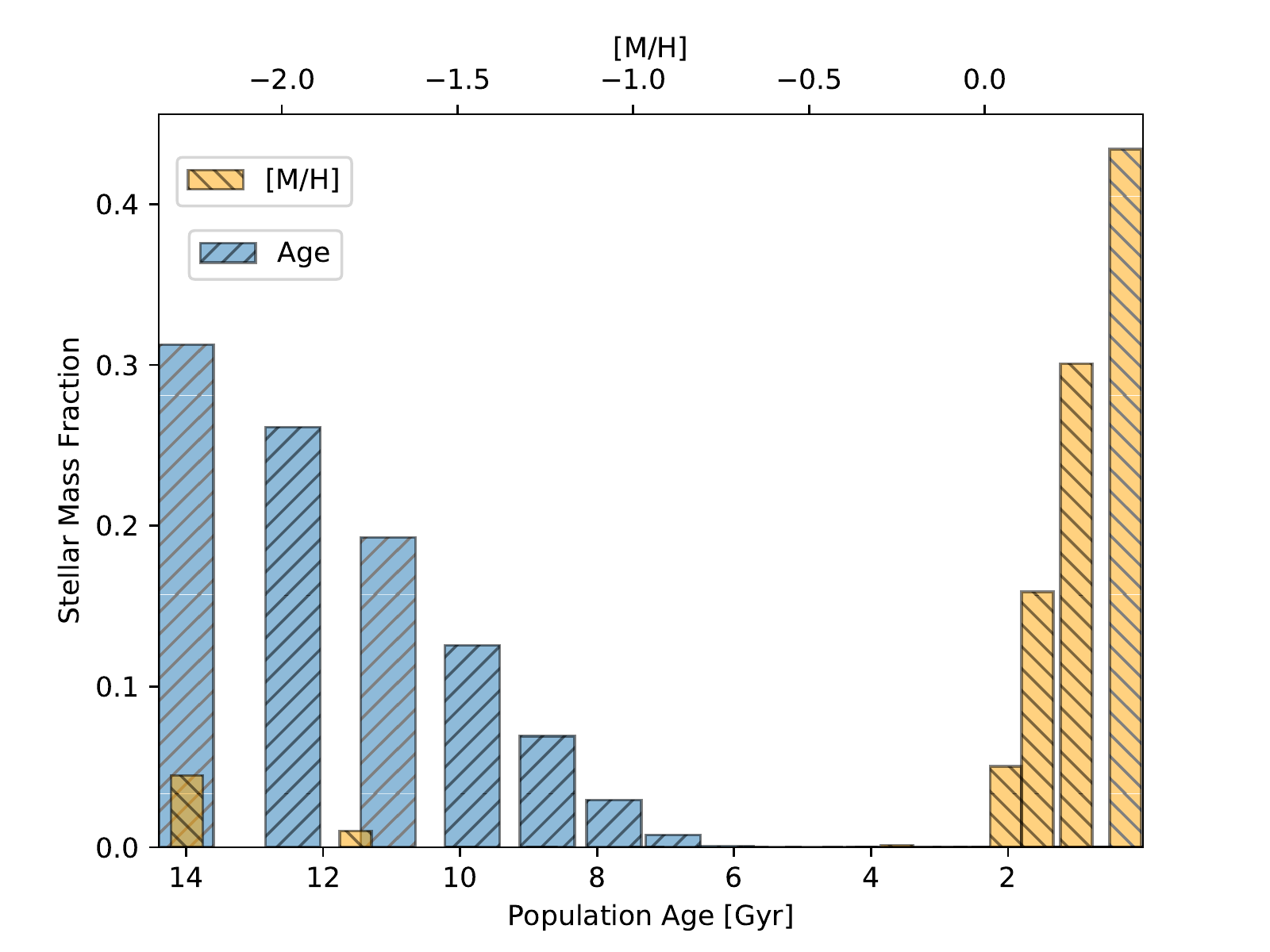}
    \caption{Star formation history and metallicity distributions as derived by using pPPXF. The bottom X-axis indicates the age of the stellar population, whereas the top one indicates the value of the metallicity ([M/H]). It can be seen from the figure that the stellar population is dominated by old and metal-rich stars.}
    \label{fig:star_formation}
\end{figure}

In a first step, we adjust the kinematics parameters, radial velocity and velocity dispersion of the stellar emission. This is performed within the range 5100-6200 \AA, in which there are several absorption features, the most prominent are Mg~b, Fe~5270, 5335 and Na-D, and it is free of contamination by emission lines. A multiplicative second order Legendre polynomial is included to adapt the template continuum shape to the observed spectrum.  The best fit, with a reduced $\chi^2\sim1$, is obtained for a velocity dispersion of 237$\pm$9\,km/s, yielding a redshift estimation of z=0.06504$\pm$0.00002. 
From the velocity dispersion we can estimate the black-hole (BH) mass using the relationship provided by \citet{McConnell11} (we use the relationship derived for the whole sample), which yields $(4.8 \pm 0.9) \times 10^8 M_{\odot}$.
Once the radial velocity and the velocity dispersion are obtained, they are kept fixed and used to determine the best fitting stellar population model. Before running pPXF over the whole spectral range, we mask the regions containing emission line features, {\it i.e.} around the H$\beta$+[OIII] and H$\alpha$+[NII] complexes. The observed spectrum resembles that of an early-type or elliptical galaxy (see Fig~\ref{fig:stellarpop}). As expected from visual inspection of the spectra, the highest weights found by pPXF of SSP models are obtained near the upper limits of age and metallicity within the available grid on the stellar library. The mass-weighted resulting parameters indicate a very old ($11.72 \pm 0.06\,\mathrm{Gyr}$) and metallic ($\mathrm{[M/H]}\simeq 0.159 \pm 0.016$) 
stellar population. The distributions of the age and metallicity of the stellar population  are shown in Fig.~\ref{fig:star_formation}. It shows that star formation has decrease with time, and the last star formation episodes took place around 6\,Gyr ago.

Note that the best fit to the optical spectrum is obtained after adding a first order Legendre polynomial (in a log($\lambda$) scale) used to improve the fit of the observed spectrum without varying the template continuum shape. The addition of this polynomial component mimics the commonly observed power law component and accounts for the likely featureless contribution from the active nucleus. As it can be seen in Fig~\ref{fig:stellarpop}, this nuclear component (featureless continuum) is very important w.r.t. the total observed optical spectrum enclosed within the slit (about 2/3) in the blue part whereas it becomes less important at the red end (about 1/10). The index of the best matching power--law is -1, {\it i.e.} $F_{\lambda} \propto \lambda^{-1}$, or equivalently $F_{\nu} \propto \nu^{-1}$.  
A check was performed to find the best description of the additive polynomial, testing different polynomial orders: for a zero-order, or simply adding a constant the $\chi^2$/d.o.f. worsen by 30\% w.r.t. the first-order polynomial, whereas for a second-order the $\chi^2$/d.o.f. improves by only 10\%, although the solution does not make physical sense because it drops rapidly towards the blue and red ends, while the featureless continuum is usually shaped as a power-law type function.

\begin{figure}
    \centering
    \includegraphics[width=0.48\textwidth]{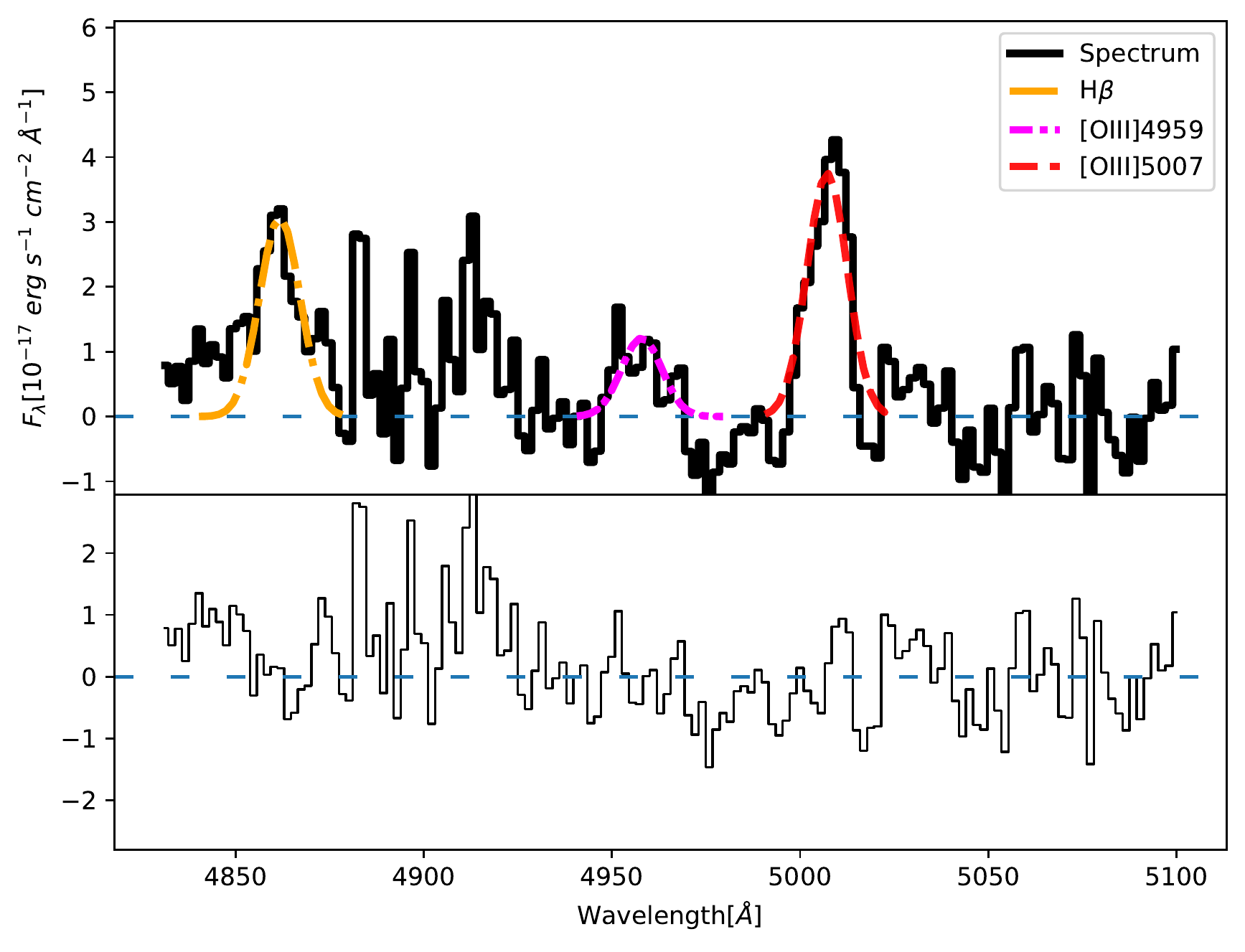}
    \includegraphics[width=0.48\textwidth]{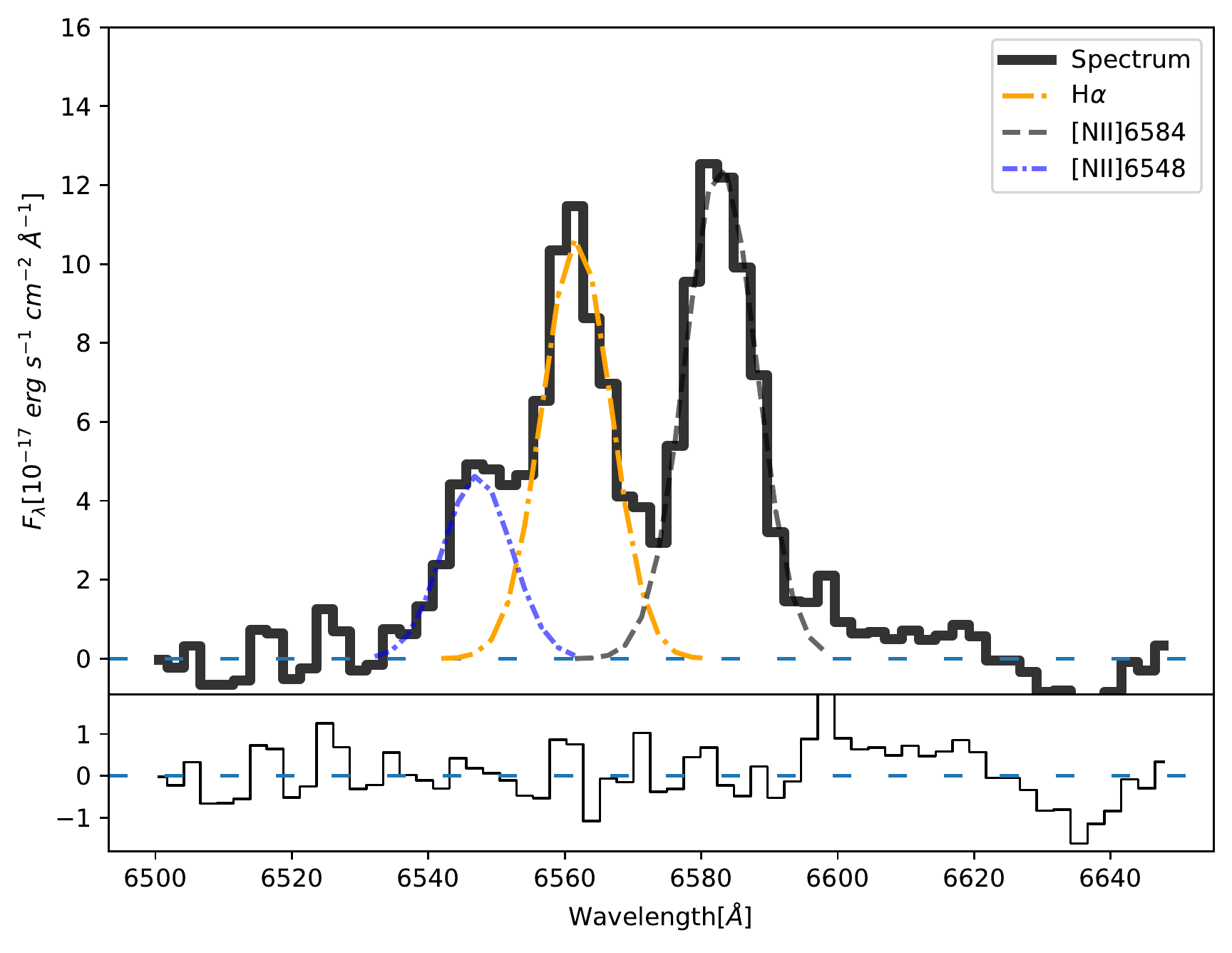}
    \caption{Gaussian fits to the emission spectral lines after subtraction of the continuum, including the residuals at the bottom panel of each figure. \emph{Top:} the fits to the H$\beta$ and [OIII] lines are represented with different colors and line styles, while the observed spectrum is denoted by the solid black line. \emph{Bottom:} the H$\alpha$+[NII]6548,6584 \AA\ complex is represented. The emission lines are fitted simultaneously, with the restriction of having the same width and the relative distance between them corresponding to the rest frame ratio. Moreover, the theoretical intensity ratio of the lines is used to constrain the relative fluxes of the [OIII] and [NII] lines. }
    \label{fig:fitHaNII}
\end{figure}

\subsection{Emission lines}
The maximum equivalent width (EW$\sim 2 \AA$) of the lines present in our spectrum correspond to the H$\alpha$+[NII] complex (see Fig.~\ref{fig:stellarpop}). According to such EW value \PGC\ is consisting with its classification as a BL Lac object.

The best stellar population model composite is subtracted from the observed spectrum in order to reveal better the ionized gas emission (see Figs. ~\ref{fig:stellarpop} and \ref{fig:fitHaNII}).  The residuals show emission line features which correspond to the H$\alpha$+[NII],  H$\beta$+[OIII] complexes and less evident to [SII]. We fit the emission features using Gaussian functions for each transition, which parameters are listed in Table~\ref{tab:emission_lines}. Due to the limited resolution power, in the case of H$\alpha$ complex we have constrained the 3 lines assuming the same velocity offset and the same line width.  
We try to classify the emission line spectrum using diagnostic diagrams as proposed by \citet{Kewley06}. After placing our measured line ratios in their diagnostic diagrams, they lie in the boundary region separating HII regions Seyfert and LINERs. Only in the [NII]/H$\alpha$ vs [OIII]/H$\beta$ our measurement lies in the right wing corresponding to AGN type.

\begin{table}[Emission Lines Measurements]
    \centering
    \begin{tabular}{lrrr}
       \hline Line ID  & Center & FWHM$\AA$ & Flux \\ 
        &   \AA  & \AA & $10^{-16}\mathrm{erg}\,\mathrm{cm}^{-2}\,\mathrm{s}^{-1}$ \\ \hline 
       H$\beta$ & $4861 \pm 4$ & $16 \pm 14$ & $3.6 \pm 1.1$ \\
       {[}OIII{]}   & $5007.9 \pm 1.2$ & $12 \pm 4$ & $5.98 \pm 0.6$ \\
       H$\alpha$  & $6562.2 \pm 0.5$ & $11.9 \pm 1.4$ &  $13.3 \pm 0.5$  \\
       {[}NII{]$^*$} & 6582.8  & 11.9 & $15.7\pm 0.5$ \\
       {[}SII{]$^{**}$} & 6715.8 & 12.8 & $2.8 \pm 0.7$ \\ 
       {[}SII{]} & $6730.8 \pm 3.7$ & $12.8 \pm 11$  &  $3.2 \pm 0.8$ \\
    \end{tabular}
    \caption{Optical emission lines from 2WHSP~J073326.7+515354 assuming a Gaussian fit. Notes: $^*$The center and width of the [NII]6548,6584 are linked to the values corresponding to H$\alpha$. The line flux ratio [NII]6584/[6548] is kept fixed to the theoretical value 3. $^{**}$The center and width of the [NII]6716,6731 are constrained among them.} 
    \label{tab:emission_lines}
\end{table}

Using the H$\alpha$ emission line flux we have derived the line luminosity, $L(\mathrm{H}\alpha)=(13.7\pm0.5)\times 10^{39}\,\mathrm{erg\,s^{-1}}$. This value is among the highest 
values found by \citet{Eracleous2010,Ho97} for a sample of LINERs and lower than the values found by \citet{Kinney91} for a sample of Seyfert 2 galaxies. From the line luminosity we derive an estimation of
the ionizing photon flux, assuming all ionizing photons (energies larger than 1 Rydberg) are absorbed by the gas emitting H$\alpha$ \citep{Osterbrock}.
Thus we obtained $Q[\mathrm{H}^0]=(10.1 \pm0.5)\times 10^{51} \mathrm{ph\,s^{-1}}$, which is among the highest values found for low luminosity AGNs  \citep{Eracleous2010}.
Using the ionizing photon flux we have estimated the corresponding flux density at frequencies close to the ionization potential of H$^0$, thus we obtained $\nu\,F_{\nu}[1Ryd]\simeq3 \times 10^{-14}\mathrm{erg}\,\mathrm{cm}^{-2}\,\mathrm{s}^{-1}$, assuming $F_{\nu} \propto \nu^{-1}$. This value results to be about 2 orders of magnitude below the measured in the W2/UVOT filter \citep{pgc_magic}. This fact suggests that the mechanism responsible of the observed UV radiation, likely synchrotron radiation from a relativistic jet, cannot be the same producing photoionization of the gas emitting H$\alpha$, unless the covering factor is very small. An alternative mechanism to produce the excitation of the gas can be due to shock-fronts driven by the radio jet \citep[see for example][]{Dopita2015}.

\section{Host galaxy morphology}
\label{sec:morphology}

\begin{figure*}
    \centering
    \includegraphics[width=0.98\textwidth]{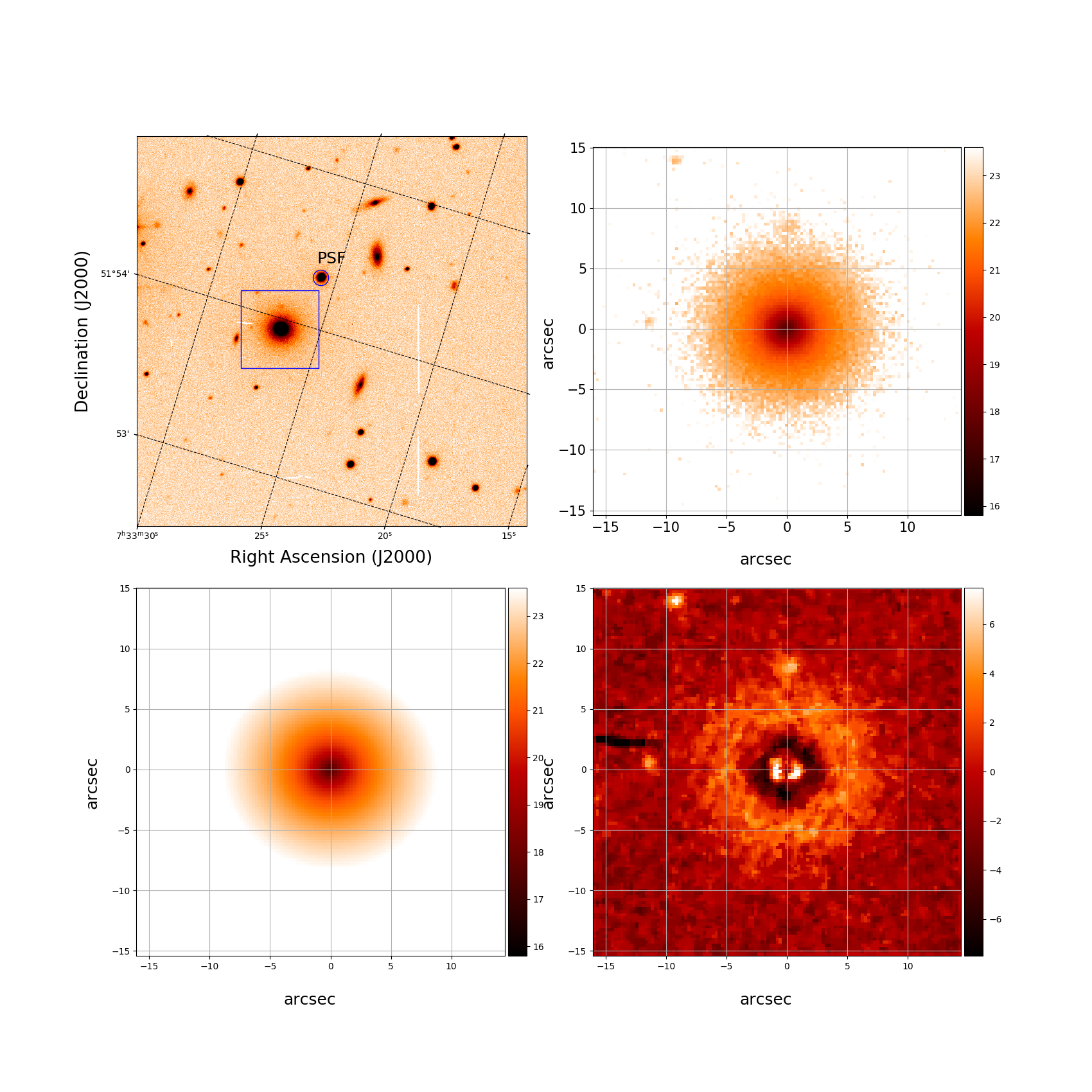}
    \caption{Left-top: Observed field with the GTC telescope, \PGC\ is marked by the blue rectangle while the star used for the PSF calculation is denoted by the blue circle. Right-top: central $15\times 15$\arcsec$^2$ of the r band image of 2WHSP~J073326.7+515354. Left-bottom: GALFIT model using a Sersic profile combined with an unresolved nuclear component. Right-bottom: Residual image after subtracting the model from the observation. Colour bars are in $mag\ arcsec^{-2}$ in the observed and model galaxy, and ADU/RMS in the residual image. North is up and East is left. }
    \label{fig:2d_galaxy_model}
\end{figure*}

We use the acquisition image taken with OSIRIS  to investigate the properties of the blazar host galaxy. Despite the short integration time (10 sec), \PGC\ appears clearly extended (see Fig.~\ref{fig:2d_galaxy_model}). This image was taken using the Sloan r filter, immediately before the spectra. The image is bias subtracted and flat-field corrected using a flat field taken from the database of the instrument.  
The photometric zero point is determined by comparing the instrumental magnitude of several stars in the field with the calibrated values from PANSTARRs catalogue.  

\begin{figure}
    \centering
    \includegraphics[width=0.48\textwidth]{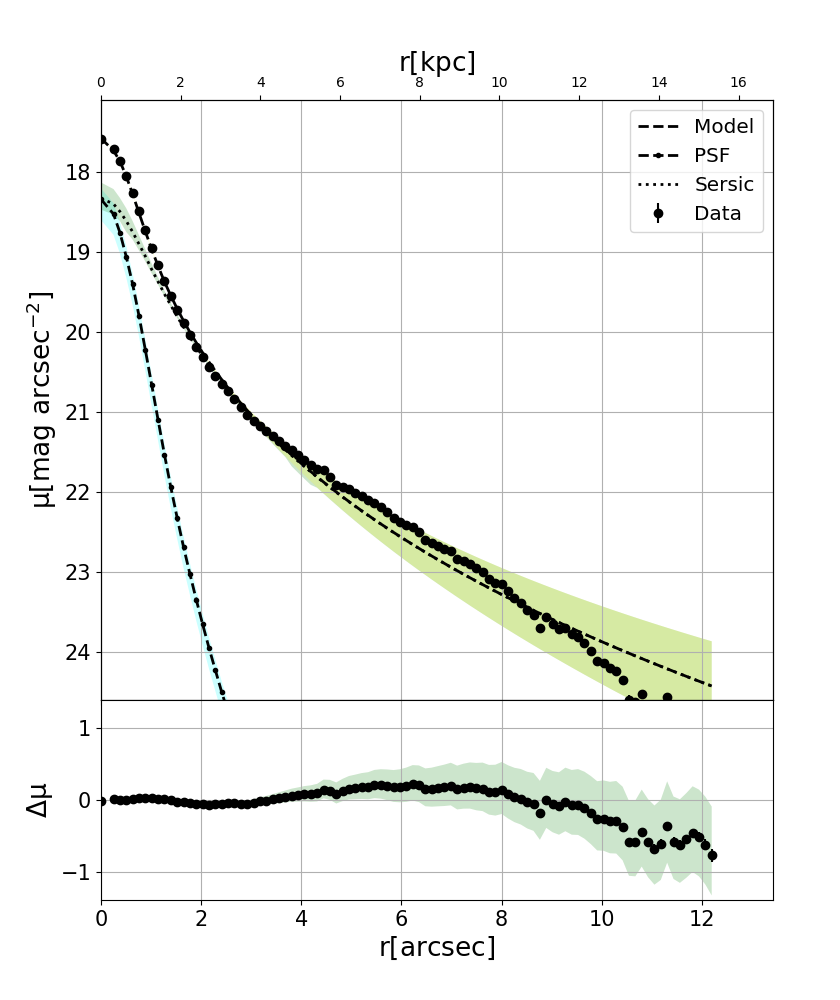}
    \caption{1D surface brightness decomposition. Observed profile is shown with black dots, the S\'ersic profile is the dotted line (green color) the scaled PSF is the dot-dashed line (blue color). Residuals are shown in the bottom panel. The shaded areas represent the results from the models after varying the background within $\pm1\sigma$.}
    \label{fig:brightness_prof}
\end{figure}

In order to study the morphology of the host galaxy, we apply the 2D surface brightness model fitting code GALFIT \citep{Peng02}, including a bulge component, characterized by a S\'ersic profile, and a point spread function (PSF) to fit the unresolved nuclear component. The 2D data and models are show in Fig.~\ref{fig:2d_galaxy_model}, and the 1D surface brightness decomposition is represented in Fig.~\ref{fig:brightness_prof}. The PSF was obtained from a nearby bright star (r=16.2), at a distance of 26 arcsec NNW from the target (marked in Fig.~\ref{fig:2d_galaxy_model}). This star was modelled using a Moffat function plus an additional Gaussian for the wings. The model was subtracted to the star image and no structure was found in the residuals.

All parameters of the S\'ersic profile and the scaling factor of the PSF model were allowed to vary. The best-fitting values are given in Table~\ref{tab:StructPar}. We estimate the parameter uncertainties by varying the sky value by $\pm1\sigma$ (corresponding to the RMS of the empty sky regions). 
The best-fitting S\'ersic index is $n=3.86$, which indicates the presence of a classical bulge, suggesting an elliptical host galaxy \citep{Blanton03, Caon93}. The ellipticity obtained from the 2D modelling is very small, corresponding to an E1 type galaxy.

The resulting model and the residuals are shown in Fig.~\ref{fig:2d_galaxy_model}. By looking at Fig.~\ref{fig:brightness_prof}, it can be noticed that the contributions of the PSF and bulge at the center are very similar, but at distance of 2" from the center, the nuclear profile is about 5 magnitudes fainter than the bulge one. The residual image shows an excess at low surface brightness in the form of a ring at distances around 5-7~\arcsec (or equivalently 6-9~kpc) from the center. Ring structures have been observed in several non-barred elliptical galaxies \citep[][ reported a double ring in PGC~1000714]{10.1093/mnras/stw3107}. The existence of such structure must be confirmed using images taken at other filters, which should give hints about its origin. Nevertheless it does not affect our results about the host morphology.

\begin{table*}[Photometric and structural parameters of the 2WHSP~J073326.7+515354 host galaxy]
    \centering
    \begin{tabular}{ccccccccc}
       \hline \multicolumn{3}{c}{PSF}  & \multicolumn{5}{c}{Sersic Model} & $\chi^2/\nu$ \\ 
      m$_r$ & M$_r$ & FWHM & m$_r$ & M$_r$ & n & $R_e$ & $b/a$ & \\  
           &    & {[}arcsec{]} & &  & & {[}arcsec/kpc{]} &  \\ \hline 
     17.97$^{18.12}_{17.87}$ & -19.51 & 0.97 & $15.40^{15.62}_{15.16}$ & -22.05 & $3.86^{5.1 7}_{3.09}$ & 3.77$^{5.66}_{2.96}$ / 4.74 &  0.92 & 0.67 \\  \hline 
    \end{tabular}
    \caption{Notes: At the distance of 2WHSP~J073326.7+515354 the angular scale is 1.257 kpc/\arcsec. The absolute magnitude has been computed using a distance modulus equals to 37.34 and corrected for insterstellar extinction (see text).}
    \label{tab:StructPar}
\end{table*}

We also compute the corresponding flux contributions of the unresolved component and the bulge integrated within the area of the aperture used in our spectroscopy. We find a contribution around 30\% for the unresolved PSF component, which is slightly above the value derived from pPXF modelling (see Section~\ref{sec:StellPop}).
We derive an integrated flux of 0.267$\pm$0.03~mJy in the Sloan r filter for this component. The derived flux is about 30\% smaller than the value reported by \citet{pgc_magic} in the R band (see their Table~A5). The two values can be compared given the small difference in effective wavelength between the two filters (around $150 \AA$). This result is expected given the fact that our PSF is much narrower than the one used in their work. 

The mass of the central black hole can be estimated from the bulge luminosity, using one of the well known M$_{BH}$--L relationships. We have used the one provided by \citet{Graham07}, which relates the absolute magnitude in R band with the black hole mass. In order to convert our measurement in the Sloan r band to R band, we use the transformation equations provided by \citet{Jordi06} which are valid for stellar spectra. Hence they can be applied to our case given the fact that the host galaxy spectrum is dominated by its stellar content. The transformation equation require an estimate of the color $(r-i)$, which is obtained from the tight relationships found by \citet{Chang06}, between absolute magnitude ($M_r$) and colours. Thus our measured absolute magnitude in the r filter ($M_r=-22.0 \pm 0.2$) converts to $M_R=-22.2 \pm 0.2$, which yields to $M_{BH} = (3.7\pm 1.0) \times 10^8\,M_{\odot}$. 
This value of $M_{BH}$ compares very well with the one estimated from the velocity dispersion of the stellar emission (see Section~\ref{sec:StellPop}). 

\section{Conclusions}
\label{sec:conclusions}

In this work we performed the first optical characterization of \PGC\ using optical spectroscopic observations with INT and GTC telescopes. Its optical spectrum is galaxy dominated as typically seen in case of EHBLs. This study resulted in the first estimation of its distance ($\mathrm{z}=0.06504\pm0.00002$) which is crucial for the gamma-ray studies in order to estimate the effect of the distortion imprinted by the interaction with the EBL. 

A comparison with the stellar population synthesis models MILES \citep{Vazdekis10} allowed us to classify the host galaxy as an elliptical galaxy. The stellar population within the galaxy is an old population reaching it maximum at $(11.72 \pm 0.06)\,\mathrm{Gyr}$ and a metallicity of $\mathrm{[M/H]}\simeq 0.159 \pm 0.016$. The acquisition image clearly display an extended target, which is used to carry out a morphological study. As shown in Fig.~\ref{fig:brightness_prof}, the radial profile can be explained using two components: a nuclear region and a classical extended bulge.

The black hole mass is derived using two different methods. On one hand, a mass of $(4.8\pm0.9)\times10^8\,M_{\odot}$ is derived using the prescription from \citet{McConnell11} for the relation with the velocity dispersion. On the other hand, a mass of $(3.7\pm1.0)\times10^8\,M_{\odot}$ is estimated when using the relation of the absolute magnitude of the bulge in R band with the black hole mass as given by \cite{Graham07}. Both estimations of the black hole mass are consistent within $\sim20\%$, and within their uncertainty estimations. These results are very well compatible with the black hole masses derived for a large sample of gamma-ray BL Lac objects reported by \cite{shaw}.


\section{Acknowledgements}

This work is based on observations made with the GTC telescope, in the Spanish Observatorio del Roque de los Muchachos of the Instituto de Astrofisica de Canarias, under Director\textquotesingle s Discretionary Time (proposal code GTC2018-099, PI: J. Becerra). JBG acknowledges the support of the Viera y Clavijo program funded by ACIISI and ULL. We thank Luca Foffano and Elisa Prandini for useful discussions about the target under study in this work.



\bibliographystyle{mnras}
\bibliography{bibliography} 

\begin{thebibliography}{}
\makeatletter
\relax
\def\mn@urlcharsother{\let\do\@makeother \do\$\do\&\do\#\do\^\do\_\do\%\do\~}
\def\mn@doi{\begingroup\mn@urlcharsother \@ifnextchar [ {\mn@doi@}
  {\mn@doi@[]}}
\def\mn@doi@[#1]#2{\def\@tempa{#1}\ifx\@tempa\@empty \href
  {http://dx.doi.org/#2} {doi:#2}\else \href {http://dx.doi.org/#2} {#1}\fi
  \endgroup}
\def\mn@eprint#1#2{\mn@eprint@#1:#2::\@nil}
\def\mn@eprint@arXiv#1{\href {http://arxiv.org/abs/#1} {{\tt arXiv:#1}}}
\def\mn@eprint@dblp#1{\href {http://dblp.uni-trier.de/rec/bibtex/#1.xml}
  {dblp:#1}}
\def\mn@eprint@#1:#2:#3:#4\@nil{\def\@tempa {#1}\def\@tempb {#2}\def\@tempc
  {#3}\ifx \@tempc \@empty \let \@tempc \@tempb \let \@tempb \@tempa \fi \ifx
  \@tempb \@empty \def\@tempb {arXiv}\fi \@ifundefined
  {mn@eprint@\@tempb}{\@tempb:\@tempc}{\expandafter \expandafter \csname
  mn@eprint@\@tempb\endcsname \expandafter{\@tempc}}}

\bibitem[\protect\citeauthoryear{{Acciari} et~al.,}{{Acciari}
  et~al.}{2019}]{ebl_magic}
{Acciari} V.~A.,  et~al., 2019, \mn@doi [\mnras] {10.1093/mnras/stz943}, \href
  {https://ui.adsabs.harvard.edu/abs/2019MNRAS.486.4233A} {486, 4233}

\bibitem[\protect\citeauthoryear{{Blanton} et~al.,}{{Blanton}
  et~al.}{2003}]{Blanton03}
{Blanton} M.~R.,  et~al., 2003, \mn@doi [\apj] {10.1086/375528}, \href
  {https://ui.adsabs.harvard.edu/abs/2003ApJ...594..186B} {594, 186}

\bibitem[\protect\citeauthoryear{{Caon}, {Capaccioli}  \& {D'Onofrio}}{{Caon}
  et~al.}{1993}]{Caon93}
{Caon} N.,  {Capaccioli} M.,   {D'Onofrio} M.,  1993, \mn@doi [\mnras]
  {10.1093/mnras/265.4.1013}, \href
  {https://ui.adsabs.harvard.edu/abs/1993MNRAS.265.1013C} {265, 1013}

\bibitem[\protect\citeauthoryear{{Cappellari}}{{Cappellari}}{2017}]{Cappellari17}
{Cappellari} M.,  2017, \mn@doi [\mnras] {10.1093/mnras/stw3020}, \href
  {https://ui.adsabs.harvard.edu/abs/2017MNRAS.466..798C} {466, 798}

\bibitem[\protect\citeauthoryear{{Cappellari} \& {Emsellem}}{{Cappellari} \&
  {Emsellem}}{2004}]{Cappellari04}
{Cappellari} M.,  {Emsellem} E.,  2004, \mn@doi [\pasp] {10.1086/381875}, \href
  {https://ui.adsabs.harvard.edu/abs/2004PASP..116..138C} {116, 138}

\bibitem[\protect\citeauthoryear{{Chang}, {Gallazzi}, {Kauffmann}, {Charlot},
  {Ivezi{\'c}}, {Brinchmann}  \& {Heckman}}{{Chang} et~al.}{2006}]{Chang06}
{Chang} R.,  {Gallazzi} A.,  {Kauffmann} G.,  {Charlot} S.,  {Ivezi{\'c}}
  {\v{Z}}.,  {Brinchmann} J.,   {Heckman} T.~M.,  2006, \mn@doi [\mnras]
  {10.1111/j.1365-2966.2005.09778.x}, \href
  {https://ui.adsabs.harvard.edu/abs/2006MNRAS.366..717C} {366, 717}

\bibitem[\protect\citeauthoryear{Chang, Arsioli, Giommi  \& Padovani}{Chang
  et~al.}{2017}]{2WHSP_cat}
Chang Y.~L.,  Arsioli B.,  Giommi P.,   Padovani P.,  2017, \mn@doi [Astron.
  Astrophys.] {10.1051/0004-6361/201629487}, 598, A17

\bibitem[\protect\citeauthoryear{{Dopita} et~al.,}{{Dopita}
  et~al.}{2015}]{Dopita2015}
{Dopita} M.~A.,  et~al., 2015, \mn@doi [\apj] {10.1088/0004-637X/801/1/42},
  \href {https://ui.adsabs.harvard.edu/abs/2015ApJ...801...42D} {801, 42}

\bibitem[\protect\citeauthoryear{{Eracleous}, {Hwang}  \& {Flohic}}{{Eracleous}
  et~al.}{2010}]{Eracleous2010}
{Eracleous} M.,  {Hwang} J.~A.,   {Flohic} H. M.~L.~G.,  2010, \mn@doi [\apj]
  {10.1088/0004-637X/711/2/796}, \href
  {https://ui.adsabs.harvard.edu/abs/2010ApJ...711..796E} {711, 796}

\bibitem[\protect\citeauthoryear{{Falc{\'o}n-Barroso},
  {S{\'a}nchez-Bl{\'a}zquez}, {Vazdekis}, {Ricciardelli}, {Cardiel}, {Cenarro},
  {Gorgas}  \& {Peletier}}{{Falc{\'o}n-Barroso} et~al.}{2011}]{Falcon11}
{Falc{\'o}n-Barroso} J.,  {S{\'a}nchez-Bl{\'a}zquez} P.,  {Vazdekis} A.,
  {Ricciardelli} E.,  {Cardiel} N.,  {Cenarro} A.~J.,  {Gorgas} J.,
  {Peletier} R.~F.,  2011, \mn@doi [\aap] {10.1051/0004-6361/201116842}, \href
  {https://ui.adsabs.harvard.edu/abs/2011A&A...532A..95F} {532, A95}

\bibitem[\protect\citeauthoryear{{Fitzpatrick} \& {Massa}}{{Fitzpatrick} \&
  {Massa}}{2007}]{Fitzpatrick07}
{Fitzpatrick} E.~L.,  {Massa} D.,  2007, \mn@doi [\apj] {10.1086/518158}, \href
  {https://ui.adsabs.harvard.edu/abs/2007ApJ...663..320F} {663, 320}

\bibitem[\protect\citeauthoryear{{Gon{\c{c}}alves} \& {Serote
  Roos}}{{Gon{\c{c}}alves} \& {Serote Roos}}{2004}]{2004A&A...413...97G}
{Gon{\c{c}}alves} A.~C.,  {Serote Roos} M.,  2004, \mn@doi [\aap]
  {10.1051/0004-6361:20031494}, \href
  {https://ui.adsabs.harvard.edu/abs/2004A&A...413...97G} {413, 97}

\bibitem[\protect\citeauthoryear{{Graham}}{{Graham}}{2007}]{Graham07}
{Graham} A.~W.,  2007, \mn@doi [\mnras] {10.1111/j.1365-2966.2007.11950.x},
  \href {https://ui.adsabs.harvard.edu/abs/2007MNRAS.379..711G} {379, 711}

\bibitem[\protect\citeauthoryear{Gullikson, Dodson-Robinson  \&
  Kraus}{Gullikson et~al.}{2014}]{Gullikson14}
Gullikson K.,  Dodson-Robinson S.,   Kraus A.,  2014, \mn@doi [The Astronomical
  Journal] {10.1088/0004-6256/148/3/53}, 148, 53

\bibitem[\protect\citeauthoryear{{Ho}, {Filippenko}  \& {Sargent}}{{Ho}
  et~al.}{1997}]{Ho97}
{Ho} L.~C.,  {Filippenko} A.~V.,   {Sargent} W. L.~W.,  1997, \mn@doi [\apjs]
  {10.1086/313041}, \href
  {https://ui.adsabs.harvard.edu/abs/1997ApJS..112..315H} {112, 315}

\bibitem[\protect\citeauthoryear{{Jordi}, {Grebel}  \& {Ammon}}{{Jordi}
  et~al.}{2006}]{Jordi06}
{Jordi} K.,  {Grebel} E.~K.,   {Ammon} K.,  2006, \mn@doi [\aap]
  {10.1051/0004-6361:20066082}, \href
  {https://ui.adsabs.harvard.edu/abs/2006A&A...460..339J} {460, 339}

\bibitem[\protect\citeauthoryear{{Kewley}, {Groves}, {Kauffmann}  \&
  {Heckman}}{{Kewley} et~al.}{2006}]{Kewley06}
{Kewley} L.~J.,  {Groves} B.,  {Kauffmann} G.,   {Heckman} T.,  2006, \mn@doi
  [\mnras] {10.1111/j.1365-2966.2006.10859.x}, \href
  {https://ui.adsabs.harvard.edu/abs/2006MNRAS.372..961K} {372, 961}

\bibitem[\protect\citeauthoryear{{Kinney}, {Antonucci}, {Ward}, {Wilson}  \&
  {Whittle}}{{Kinney} et~al.}{1991}]{Kinney91}
{Kinney} A.~L.,  {Antonucci} R.~R.~J.,  {Ward} M.~J.,  {Wilson} A.~S.,
  {Whittle} M.,  1991, \mn@doi [\apj] {10.1086/170339}, \href
  {https://ui.adsabs.harvard.edu/abs/1991ApJ...377..100K} {377, 100}

\bibitem[\protect\citeauthoryear{{Kroupa}}{{Kroupa}}{2001}]{Kroupa01}
{Kroupa} P.,  2001, \mn@doi [\mnras] {10.1046/j.1365-8711.2001.04022.x}, \href
  {https://ui.adsabs.harvard.edu/abs/2001MNRAS.322..231K} {322, 231}

\bibitem[\protect\citeauthoryear{{MAGIC Collaboration} et~al.,}{{MAGIC
  Collaboration} et~al.}{2019}]{pgc_magic}
{MAGIC Collaboration} et~al., 2019, \mn@doi [\mnras] {10.1093/mnras/stz2725},
  \href {https://ui.adsabs.harvard.edu/abs/2019MNRAS.490.2284M} {490, 2284}

\bibitem[\protect\citeauthoryear{{McConnell}, {Ma}, {Gebhardt}, {Wright},
  {Murphy}, {Lauer}, {Graham}  \& {Richstone}}{{McConnell}
  et~al.}{2011}]{McConnell11}
{McConnell} N.~J.,  {Ma} C.-P.,  {Gebhardt} K.,  {Wright} S.~A.,  {Murphy}
  J.~D.,  {Lauer} T.~R.,  {Graham} J.~R.,   {Richstone} D.~O.,  2011, \mn@doi
  [\nat] {10.1038/nature10636}, \href
  {https://ui.adsabs.harvard.edu/abs/2011Natur.480..215M} {480, 215}

\bibitem[\protect\citeauthoryear{Mutlu~Pakdil, Mangedarage, Seigar  \&
  Treuthardt}{Mutlu~Pakdil et~al.}{2016}]{10.1093/mnras/stw3107}
Mutlu~Pakdil B.,  Mangedarage M.,  Seigar M.~S.,   Treuthardt P.,  2016,
  \mn@doi [Monthly Notices of the Royal Astronomical Society]
  {10.1093/mnras/stw3107}, 466, 355

\bibitem[\protect\citeauthoryear{{Osterbrock}}{{Osterbrock}}{1989}]{Osterbrock}
{Osterbrock} D.~E.,  1989, University Science Books, ISBN 0-935702-22-9

\bibitem[\protect\citeauthoryear{{Peng}, {Ho}, {Impey}  \& {Rix}}{{Peng}
  et~al.}{2002}]{Peng02}
{Peng} C.~Y.,  {Ho} L.~C.,  {Impey} C.~D.,   {Rix} H.-W.,  2002, \mn@doi [\aj]
  {10.1086/340952}, \href
  {https://ui.adsabs.harvard.edu/abs/2002AJ....124..266P} {124, 266}

\bibitem[\protect\citeauthoryear{{Pietrinferni}, {Cassisi}, {Salaris}  \&
  {Hidalgo}}{{Pietrinferni} et~al.}{2013}]{Pietrinferni13}
{Pietrinferni} A.,  {Cassisi} S.,  {Salaris} M.,   {Hidalgo} S.,  2013, \mn@doi
  [\aap] {10.1051/0004-6361/201321950}, \href
  {https://ui.adsabs.harvard.edu/abs/2013A&A...558A..46P} {558, A46}

\bibitem[\protect\citeauthoryear{{S{\'a}nchez-Bl{\'a}zquez}
  et~al.,}{{S{\'a}nchez-Bl{\'a}zquez} et~al.}{2006}]{SanchezBlazquez06}
{S{\'a}nchez-Bl{\'a}zquez} P.,  et~al., 2006, \mn@doi [\mnras]
  {10.1111/j.1365-2966.2006.10699.x}, \href
  {https://ui.adsabs.harvard.edu/abs/2006MNRAS.371..703S} {371, 703}

\bibitem[\protect\citeauthoryear{{Schlafly} \& {Finkbeiner}}{{Schlafly} \&
  {Finkbeiner}}{2011}]{Schlafly11}
{Schlafly} E.~F.,  {Finkbeiner} D.~P.,  2011, \mn@doi [\apj]
  {10.1088/0004-637X/737/2/103}, \href
  {https://ui.adsabs.harvard.edu/abs/2011ApJ...737..103S} {737, 103}

\bibitem[\protect\citeauthoryear{{Serote Roos} \& {Gon{\c{c}}alves}}{{Serote
  Roos} \& {Gon{\c{c}}alves}}{2004}]{2004A&A...413...91S}
{Serote Roos} M.,  {Gon{\c{c}}alves} A.~C.,  2004, \mn@doi [\aap]
  {10.1051/0004-6361:20031493}, \href
  {https://ui.adsabs.harvard.edu/abs/2004A&A...413...91S} {413, 91}

\bibitem[\protect\citeauthoryear{{Shaw} et~al.,}{{Shaw} et~al.}{2013}]{shaw}
{Shaw} M.~S.,  et~al., 2013, \mn@doi [\apj] {10.1088/0004-637X/764/2/135},
  \href {https://ui.adsabs.harvard.edu/abs/2013ApJ...764..135S} {764, 135}

\bibitem[\protect\citeauthoryear{{The Fermi-LAT collaboration}}{{The Fermi-LAT
  collaboration}}{2019}]{4fgl}
{The Fermi-LAT collaboration} 2019, arXiv e-prints, \href
  {https://ui.adsabs.harvard.edu/abs/2019arXiv190210045T} {p. arXiv:1902.10045}

\bibitem[\protect\citeauthoryear{{Vazdekis}, {S{\'a}nchez-Bl{\'a}zquez},
  {Falc{\'o}n-Barroso}, {Cenarro}, {Beasley}, {Cardiel}, {Gorgas}  \&
  {Peletier}}{{Vazdekis} et~al.}{2010}]{Vazdekis10}
{Vazdekis} A.,  {S{\'a}nchez-Bl{\'a}zquez} P.,  {Falc{\'o}n-Barroso} J.,
  {Cenarro} A.~J.,  {Beasley} M.~A.,  {Cardiel} N.,  {Gorgas} J.,   {Peletier}
  R.~F.,  2010, \mn@doi [\mnras] {10.1111/j.1365-2966.2010.16407.x}, \href
  {https://ui.adsabs.harvard.edu/abs/2010MNRAS.404.1639V} {404, 1639}

\makeatother
\end{thebibliography}

\bsp	
\label{lastpage}
\end{document}